\documentclass[fleqn,usenatbib]{mnras}

\usepackage{newtxtext,newtxmath}
\usepackage[T1]{fontenc}

\DeclareRobustCommand{\VAN}[3]{#2}
\let\VANthebibliography\thebibliography
\def\thebibliography{\DeclareRobustCommand{\VAN}[3]{##3}\VANthebibliography}

\usepackage{graphicx}	% Including figure files
\usepackage{amsmath}	% Advanced maths commands
\usepackage{xcolor}     % text color
\usepackage{tikz}
\usepackage{hyperref}
\usepackage{times}
\usepackage[normalem]{ulem}

\usepackage{etoolbox}
\makeatletter

\definecolor{myblue}{rgb}{0.24, 0.27, 0.79}

\definecolor{mygreen}{rgb}{0.00, 0.0, 0.1}
\newcommand{\textJoaquin}[1]{\textcolor{black}{#1}}
\definecolor{textRed}{rgb}{0.0, 0.00, 0.01}
\newcommand{\textRed}[1]{\textcolor{textRed}{#1}}
\title[Utilising sub-resolution halos]{Making use of sub-resolution halos in N-body simulations}

\author[J. Armijo et al.]{
Joaquin Armijo,$^{1,2}$\thanks{E-mail: joaquin.a.armijo-torres@durham.ac.uk}
Carlton M. Baugh,$^{1,2}$
Nelson D. Padilla,$^{4,5}$
Peder Norberg$^{1,3}$
and Christian Arnold$^{1}$
\\
$^{1}$Institute for Computational Cosmology, Department of Physics, Durham University, South Road, Durham, DH1 3LE, UK\\
$^{2}$Institute for Data Science, Durham University, South Road, Durham, DH1 3LE, UK\\
$^{3}$Centre for Extragalactic Astronomy, Department of Physics, Durham University, South Road, Durham DH1 3LE, UK\\
$^{4}$Instituto de Astrofísica, Pontificia Universidad Cat\'olica de Chile, Vicuña Mackenna 4860, Santiago, Chile\\
$^{5}$ Centro de Astro-Ingeniería, Pontificia Universidad Cat\'olica de Chile, Vicuña Mackenna 4860, Santiago, Chile
}
\date{Accepted XXX. Received YYY; in original form ZZZ}

\pubyear{2021}

\begin{document}
\label{firstpage}
\pagerange{\pageref{firstpage}--\pageref{lastpage}}
\maketitle

\begin{abstract}
Conservative mass limits are often imposed on the dark matter halo catalogues extracted from N-body simulations. By comparing simulations with different mass resolutions, \textRed{at $z=0$} we find that even for halos resolved by 100 particles, the lower resolution simulation predicts a cumulative halo abundance that is 5 per cent lower than in the higher resolution simulation. We propose a simple weighting scheme to utilise the halos that are usually regarded as being `sub-resolution'. With the scheme, we are able to use halos which contain only 11 particles to reproduce the clustering measured in the higher resolution simulation to within 5 per cent on scales down to $2 h^{-1}$ Mpc, thereby extending the useful halo resolution by a factor of ten below the mass at which the mass functions in the two simulations first start to deviate. \textRed{The performance of the method is slightly worse at higher redshift.} Our method allows a simulation to be used to probe a wider parameter space in clustering studies, for example, in a halo occupation distribution analysis. This reduces the cost of generating many simulations to estimate the covariance matrix on measurements or using a larger volume simulation to make large-scale clustering predictions.
\end{abstract}


%%%%%%%%%%%%%%%%%%%%%%%%%%%%%%%%%%%%%%%%%%%%%%%%%%

%%%%%%%%%%%%%%%%% BODY OF PAPER %%%%%%%%%%%%%%%%%%

\section{Introduction}

The mass resolution limit of dark matter halo catalogues extracted from N-body simulations is often set to satisfy a range of requirements and, as a result, can appear unnecessarily conservative for some applications. The measurement of the internal properties of halos is challenging and requires that objects are resolved by several hundred particles. For example, \cite{Bett:2007} demonstrated, using the Millennium simulation of \cite{Springel:2005}, that at least 300 particles are needed to measure halo spin robustly. On the other hand, many authors have used the same simulation to build semi-analytical galaxy formation models retaining halos down to 20 particles (e.g. \citealt{Croton:2006}), extending the mass resolution of the halo catalogue by more than an order of magnitude for this purpose, compared with that used to measure halo spin. 

Here we revisit how the mass resolution limit of a dark matter halo catalogue is set for use in a simple clustering study. The application in this case is to use the halos to build a galaxy catalogue, for example using a halo occupation distribution model (HOD) or a semi-analytical galaxy formation model to populate the halos with galaxies. The resulting `mock' galaxy catalogue will be compared to an observed sample, with the criteria for success being that the mock reproduces the abundance and clustering of the target sample to within some tolerance. Typical galaxy samples occupy a broad range of halo masses. If we impose an unduly restrictive mass limit on the halo catalogue that can be used from a simulation, this could result in the simulation not being suitable to probe a wide range of the parameter space in the HOD or SAM for a given galaxy selection. We judge the halo catalogue to be useful if it can be employed to reproduce the abundance and clustering of halos that would be measured in a higher resolution simulation; we show that this can be achieved for halos that are made up of a perhaps surprisingly low number of particles by employing a simple weighting scheme.  

Here we address two issues relating to the use of simulated halos in clustering studies. The first is to devise a robust and reproducible way to determine the mass resolution limit of a halo catalogue extracted from an N-body simulation for a clustering study. The second is to see if we can still use the halos below this resolution limit in a clustering analysis, which, as we shall see, represent a fraction or subset of the true population of halos at these masses. As these halos are deemed to be below the mass resolution limit we have set, these `sub-resolution' haloes will be treated in a different way to the resolved halos.  We will show that considering the sub-resolution halos allows us to extend the useful dynamic range of the simulation by a factor of $10$ below the formal resolution limit, so long as we are willing to tolerate some error in the clustering predictions. We describe  our clustering analysis as simple since we do not consider secondary contributions to halo clustering besides mass; the halo resolution needed to use internal halo properties to build assembly bias into mock catalogues has been discussed by  \cite{Ramakrishnan:2021}.

This letter is set out as follows: We present the simulations used for this study in Section~\ref{sec:methods:Nbody}. We review the halo mass function for different resolution simulations in Section~\ref{sec:methods:HMF}. The method to use sub-resolution haloes in the clustering analysis is explained in Section~\ref{sec:methods:weights}. We draw our conclusions in Section~\ref{sec:conclusions}.

\section{The N-body simulations} \label{sec:methods:Nbody}
\textJoaquin{We use three simulations of the standard cold dark matter cosmology with different mass resolutions. We mainly focus on two simulations from \cite{Arnold:2019}, but also consider the halo mass function from the P-Millennium \citep{Baugh2019}. The simulations from Arnold et~al. each use $2048^3$ collisionless particles in cubic boxes of length $L_{\rm{box}} = 768 h^{-1}\, \text{Mpc}$ and $1536 h^{-1}\, \text{Mpc}$, resulting in particle masses of $M_{\text{p}} = 4.9 \times 10^{9}$ and $3.6 \times 10^{10} h^{-1} M_{\odot}$, respectively. Both  simulations use the Planck cosmological parameters  \citep{Planck:2016}: $h = 0.6774$, $\Omega_{\rm m} = 0.3089$, $\Omega_{\Lambda} = 0.6911$, $\Omega_{\rm b} = 0.0486$, $\sigma_8 = 0.8159$, and $n_{\rm s} = 0.9667$. We use the simulation outputs at redshift $z=0$. 
%More details on the different snapshots and types of gravity available (GR or $f(R)$) can be found in the main paper.
} The P-Millennium run uses very similar but slightly different cosmological parameters to the above (e.g. $\Omega_{\rm M} = 0.307$; see Table~1 of \citealt{Baugh2019}). The simulation box size in this case is $L_{\rm box} = 542.16 h^{-1}$Mpc with the dark matter traced by $5040^3$ particles, resulting in a particle mass of $1.08 \times 10^{8} h^{-1} M_{\odot}$. The simulations were run with slightly different versions of the {\sc gadget} code (for the most recent description see \citealt{Springel:2020}). We henceforth refer to the Arnold et~al. runs by their box lengths, as L1536, and L768. The L1536 and L768 runs form a sequence in mass resolution completed by the P-Millennium which has the best mass resolution.  

Haloes are identified using  \textsc{subfind} \citep{Springel:2001}. The first step in this algorithm is to run the friends-of-friends (FoF) percolation scheme on the simulation particles. We set the minimum number of particles per group to be retained after the FoF step to be 20. \textsc{subfind} then finds local density maxima in the FoF particle groups, and checks to see if these structures are gravitationally bound; these objects are called subhalos. Particles that are not gravitationally bound to the subhalo are removed from its membership list. The mass of the subhalo is obtained using the spherical overdensity (SO) method \citep{ColeLacey:1996}. The SO method is applied to the gravitationally bound particles in the subhalo to find the radius within which the average density is 200 times the critical density of the universe. The halo mass, $M_{200c}$, is the sum of the particle masses within this radius. This results in some subhalos having masses with $M_{200c}<20M_{\text{p}}$, because small groups tend to be ellipsoidal in shape rather than spherical.  
%We call these the sub-resolution haloes, and we keep them on the final sample as they will be used in our analysis. 
We consider halo samples composed of main subhalos, i.e. the most massive subhalo within each FoF group.   
%\textbf{CMB: Are we using FoF halos, or main sub halos? When is the gravitational binding checked? I thought the sequence ran: FoF, subfind, check the gravitational binding, estimate the main subhalo mass -- is SO applied to the FoF particles or to the main subhalo particles? This is worth clarifying, as it is the difference between looking at gravitationally bound structures or a list of percolated particles.}

%Where does the 32 particles come from? Is the parameter just set to 32 in the code release or is there a justification? The Millennium retained subhalos to 20 particles from subfind. 
%We don't want to say that 100 particles is our limit -- we want to evalute this -- what is Christian's nominal resoltuion limit?: 32

%sim details - particle mass -- nominal halo resolution claimed -- box size -- stress clustering measurements robust out to large scales. 

%group finding - consider main halos - what is the mass defintion - how is this obtained from the number of particles returned by the FOF group finder? Is there any check that the particles are gravitationally bound to the halo? Any unbinding or it is simply the FOF particle list? 

\begin{figure}
    \centering
	\includegraphics[width=0.9\columnwidth]{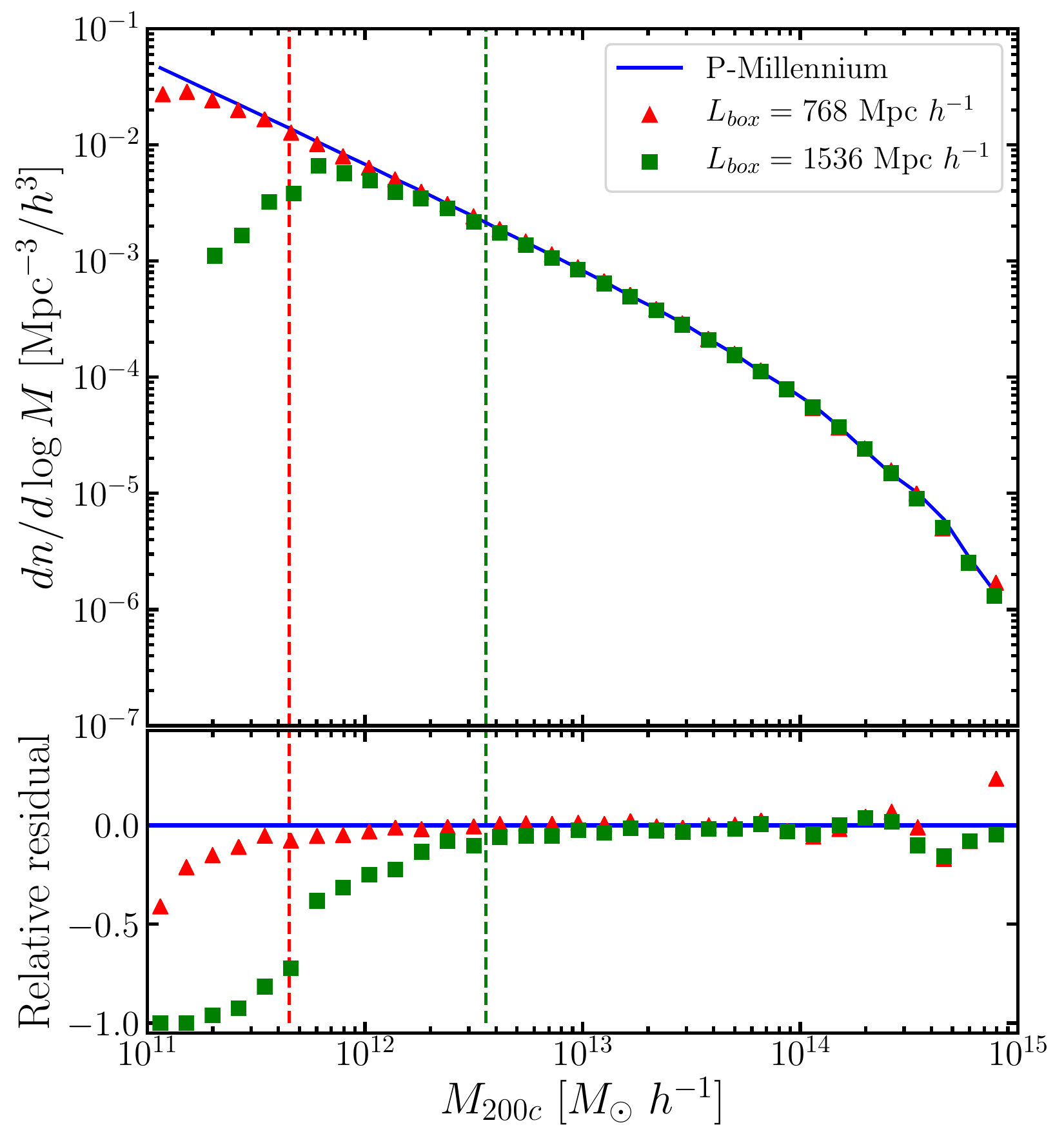}
    %\caption{The differential halo mass function at $z=0$. Top: differential mass functions, showing the fits from \cite{Tinker:2008} (solid line), \cite{Bocquet:2016} (dashed line) and the measurements from the GR $N$-body simulations of \cite{Arnold:2019} (points); red triangles show the mass function in the  $768\ \text{Mpc}\ h^{-1}$ simulation and the green squares show the $1536\ \text{Mpc}\ h^{-1}$ simulation mass function. The vertical dashed lines indicate a halo mass equivalent to 100 particles limit for the higher (red) and lower (green) resolution runs.       Bottom: fractional difference expressed relative to \cite{Tinker:2008} fit.}
    %Same as plots you have now, without the ECOSMOG points.  Plot says Tinker 2008 - should this be Tinker 2010? For papers with more than one author use Author + in the label.ie Tinker + (2010). }
    \caption{The differential halo mass function at $z=0$. Top: results from the P-Millennium  \citep{Baugh2019} (blue line), and the $\Lambda$CDM $N$-body simulations of \citealt{Arnold:2019} (points); red triangles show the mass function measured from the  L768 simulation and the green squares show the L1536 run. The vertical dashed lines indicate a halo mass of 100 particles for the L768 (red) and L1536 (green) resolution runs. Bottom: fractional difference expressed relative to the  P-Millennium halo mass function. A small correction has been applied to the masses in the P-Millennium mass function to account for the slightly different cosmological parameters used in this run and in Arnold et~al. (see text for details).}
    \label{fig1:HMF}
\end{figure}

\begin{figure*}
    \centering
%    \begin{tabular}{ccc}
%        \includegraphics[scale=0.4]{figures_pdf/2PCF_MG_GADGET_L768_1536_weigthed_logMmin_13.00_sigma0.0.pdf}&
%        \includegraphics[scale=0.4]{figures_pdf/2PCF_MG_GADGET_L768_1536_weigthed_logMmin_12.00_sigma0.0.pdf}&
	    %\includegraphics[scale=0.4]{figures_pdf/2PCF_MG_GADGET_L768_1536_weigthed_logMmin_11.60_sigma0.0.pdf}
        \includegraphics[scale=0.4]{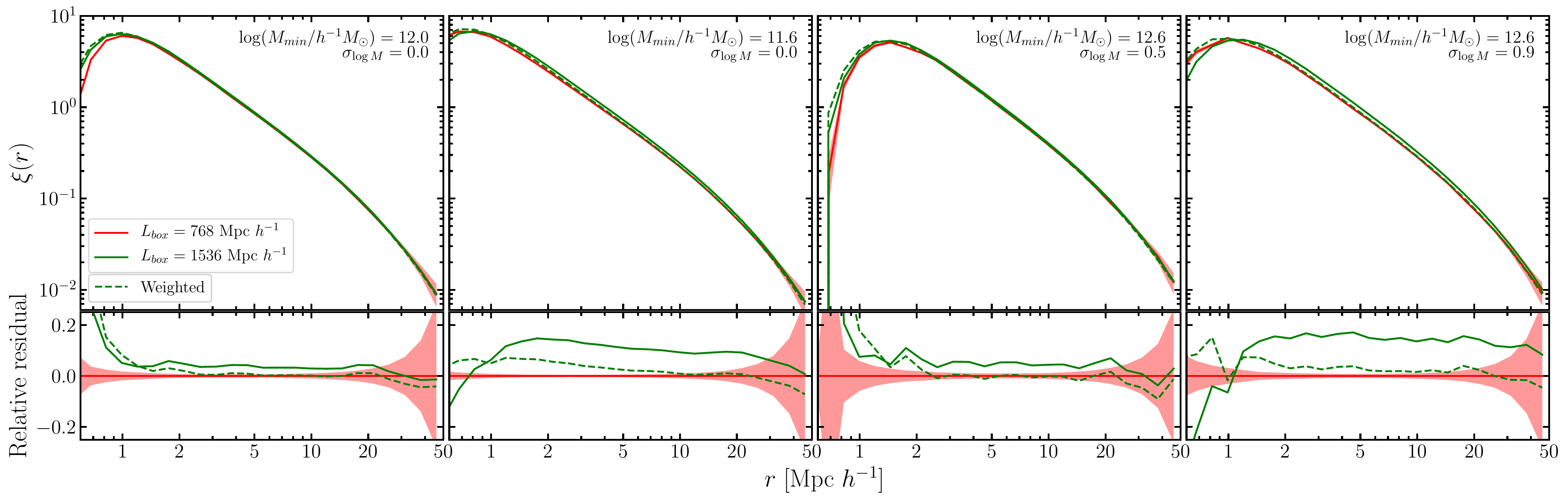}   
	%\end{tabular}
        \caption{The correlation function measured in the HR (red) and LR (green) runs for subhalo samples defined by sharp lower mass cut (left and centre-left panels, corresponding to $\sigma_{\log M} = 0$) and by a HOD-style, more gradual mass cut (centre-right and right panels, defined by  $\sigma_{\log M}>0$; see Eqn.~1).  
        For the correlation functions measured from the LR run, the solid lines shows the unweighted estimate and the dashed lines the weighted case. The lower panels show the fraction difference in the correlation function, relative to the HR measurement. The pink shading shows the error on the correlation function estimated by jackknife resampling.}
    \label{fig2:2PCF}
\end{figure*}

\section{The halo mass function and simulation resolution} \label{sec:methods:HMF}

We now look at the considerations that go into setting the mass resolution of the \textsc{subfind} halo catalogues, by  comparing the main subhalo mass functions measured in the different resolution simulations. 

Fig.~\ref{fig1:HMF} compares the mass functions measured from the L1536 and L768 simulations from Arnold et~al., with that obtained from the P-Millennium. To account for the very slightly different cosmology used in the P-Millennium, we generated analytic mass functions for the cosmologies used by Arnold et~al. and Baugh et~al. These analytic mass function are offset, and can be reconciled by applying a constant rescaling to the P-Millennium halo masses. After this correction, the differential mass functions measured from the three simulations agree with one another very well at high masses (i.e. for masses above a few times $10^{13} h^{-1} M_{\odot}$), with some fluctuations at very high halo masses which arise due to sample variance. The lower panel of Fig.~\ref{fig1:HMF} shows the fractional difference of the mass functions with respect to that measured from the P-Millennium. 
%The simulation measurements do not agree in detail with the Tinker et al fit, being around 20 per cent lower around $10^{13} h^{-1} M_{\odot}$. This deficit increases weakly on moving to lower masses. Tinker et~al. fitted to the halo mass functions measured from a range of simulation box sizes and mass resolutions, finding a scatter of 5 per cent around their fit. Also, they considered main halos rather than main subhalos.  Our measurements of the mass function are somewhat closer to the more recent fit to the halo mass function presented by \cite{Bocquet:2016}. 
The scheme we set out below depends on the comparison between the halo mass functions from the L1536 and L768 runs. 

The green vertical dashed line in Fig.~\ref{fig1:HMF} shows a halo mass corresponding to 100 particles in the L1536 simulation, i.e. $3.6 \times 10^{12} h^{-1} M_{\odot}$. At this mass, there is already a clear difference in the mass functions measured from the two simulation boxes. To quantify these differences, above a mass threshold of $10^{13} h^{-1} M_{\odot}$ there is already a 3 per cent deficit in the cumulative abundance of halos in the lower resolution L1536 run compared with the higher resolution L768 one; this rises to 12 per cent for a mass threshold of  $10^{12} h^{-1} M_{\odot}$ and 32 per cent for a mass limit of $4 \times 10^{11} h^{-1} M_{\odot}$. We have checked that the difference in the slope of the mass function between the L1536 and L768 runs is due to the difference in mass  resolution rather than sample variance in the smaller volume higher resolution box by measuring the mass functions from the larger volume simulation after splitting it into eight smaller subvolumes, each equal in volume to that of the L768 run. We found that there is remarkably little variation in the slope of the mass function around $10^{13} h^{-1} M_{\odot}$ due to sample variance. 

Fig.~\ref{fig1:HMF} shows that moving to masses below 100 particles in L1536, there is a sudden drop in the number of halos recovered in the L1536 run compared to the L768 run around $10^{12} h^{-1} M_{\odot}$. The red vertical dashed line is equivalent to 100 particles in the L768 run. The question of determining the halo mass resolution of the simulation can therefore be framed in terms of the tolerance for errors in the statistic of interest. If the halo mass function is of primary interest, then if we treat the L768 simulation as the reference or `gold standard', we could choose the resolution limit of the L1536 run as being 100 particles, in the knowledge that this gives us a $\sim$5 per cent underestimate of the cumulative abundance of dark matter halos compared to the L768 run; if we require a better reproduction of the cumulative halo abundance, then we would need to apply a mass limit greater than 100 particles. If our interest in the halos is broader and extends to clustering then we also need to assess the errors made in statistics such as the two point correlation function. It is possible, however, as we demonstrate in the next section, to extend the useful resolution of the simulation by applying a simple weighting scheme to the halos when computing their abundance and clustering. 

%\begin{figure*}
%    \centering
    %\begin{tabular}{ccc}
    %    \includegraphics[scale=0.4]{figures_pdf/2PCF_MG_GADGET_L768_1536_weigthed_logMmin_12.60_sigma0.1.pdf}&
    %    \includegraphics[scale=0.4]{figures_pdf/2PCF_MG_GADGET_L768_1536_weigthed_logMmin_12.60_sigma0.5.pdf}&
	%    \includegraphics[scale=0.4]{figures_pdf/2PCF_MG_GADGET_L768_1536_weigthed_logMmin_12.60_sigma0.9.pdf}
        %\includegraphics[scale=0.45]{figures_pdf/2PCF_MG_GADGET_L768_1536_weigthed_logMmin_12.60_sigma_0.1_0.5_0.9.pdf}
	%\end{tabular}
        %\caption{The correlation function measured in the HR (red) and LR (green) runs for subhalo samples defined by a gradual lower mass cut, in which the width of the transition from a mean occupancy of zero to one central galaxy per main subhalo is controlled by the parameter $\sigma_{\log M}$. Each column shows the results for a different value of $\sigma_{\log M}$ for $M_{\rm min} = 10^{12.6} h^{-1} M_{\odot}$: $\sigma_{\log M}=0.1$ (left), $\sigma_{\log M}=0.5$ (middle) and $\sigma_{\log M}=0.9$ (right). For the correlation functions measured from the LR run, the solid lines shows the unweighted estimate and the dashed lines the weighted case. The lower panels show the fraction difference in the correlation function, relative to the HR measurement. The pink shading shows the error on the correlation function estimated by jackknife resampling.  \textbf{CMB - fix labels: $\log (M_{min}/(h^{-1}M_{\odot})) = 13.0$ and $\sigma_{\log M} = 0.0$ (to match the notation used in text)
         %}
        %}
    %\label{fig2:2PCFb}
%\end{figure*}

\begin{figure*}
    \centering
    \begin{tabular}{ccc}
	\includegraphics[scale=0.33]{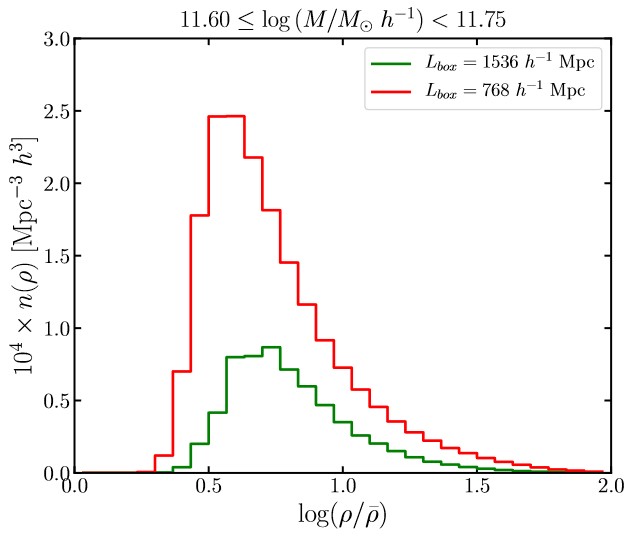}&
	\includegraphics[scale=0.33]{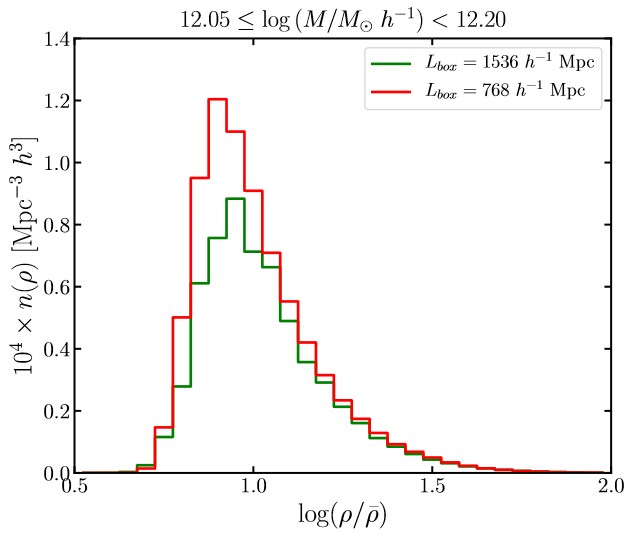}&
	\includegraphics[scale=0.33]{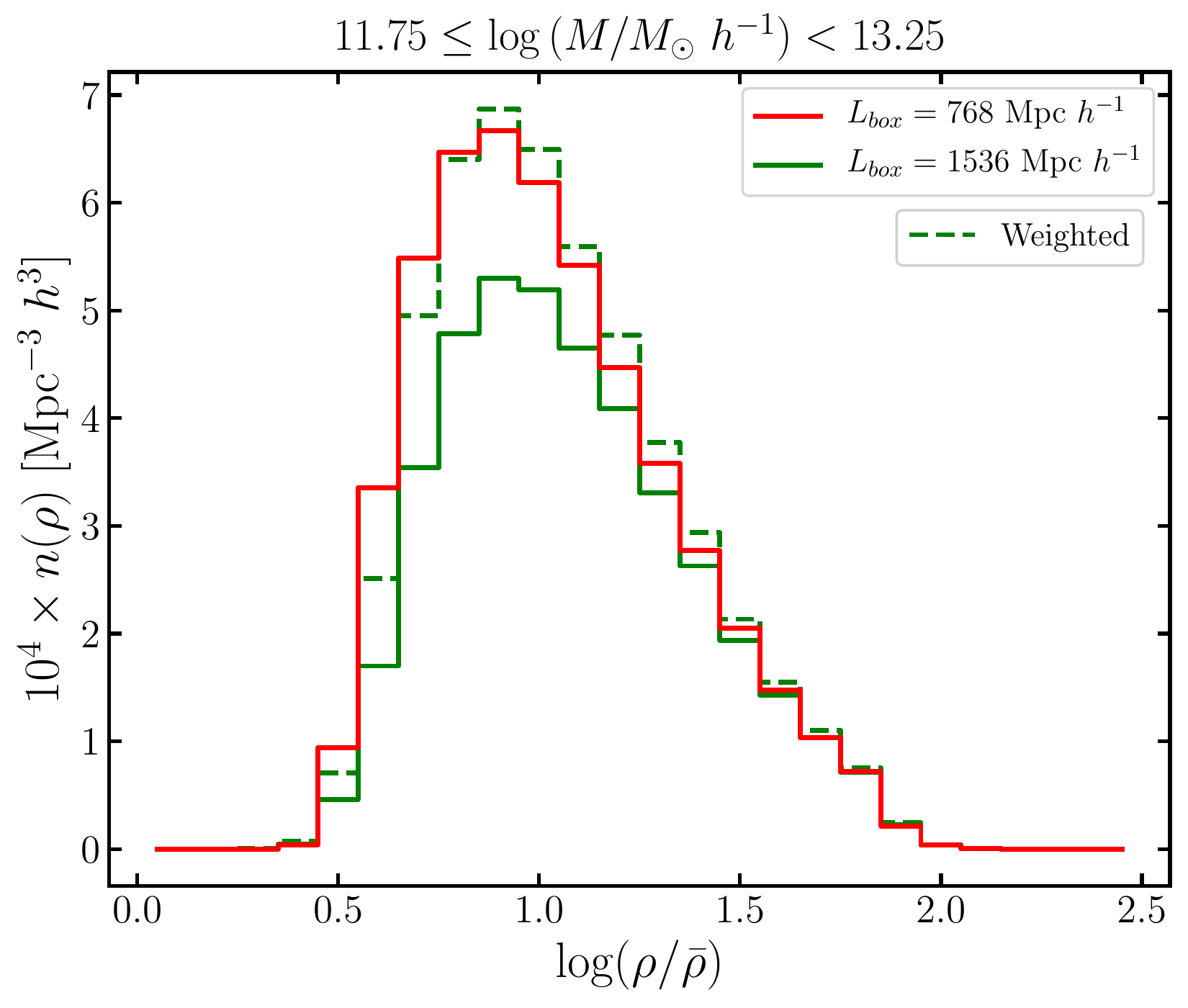}
	\end{tabular}
    \caption{The distribution of matter density counts in cells of size $1.6\ h^{-1}\ \text{Mpc}$ centred on halos in the stated mass range, measured from the L768 (red) and L1536 (green) simulations. The difference in volume of the L1536 and L768 runs has been taken into account in the normalisation. The left and central panels show the count in cells distributions for the bins used in the mass function (the bin limits are written at the top of each panel) for which the weights are greater than unity. The right panel shows the distribution of cells for a wider mass range covering all of the bins for which  the weights in our scheme are greater than unity. Here the green dashed line shows the distribution of counts-in-cells in the L768 simulation after applying the weights. }
    \label{fig3:density}
\end{figure*}

\section{Extending the resolution of the simulated halo catalogue} \label{sec:methods:weights}

Typically, a clustering study involves using the number density of objects and the two-point correlation function estimated from an observational sample to constrain a model, such as setting the parameter values in an HOD model. Here, we present a simple weighting scheme that extends the resolution limit of a simulated halo catalogue down to lower halo masses than are generally considered for use in clustering analyses. The scheme returns the exact abundance of clustering tracers, by construction, and yields a more accurate prediction of the two-point clustering to within some tolerance. The procedure used to derive the resolution limit is transparent and reproducible. 

The weighting scheme is remarkably simple. A weight is defined in bins of halo mass such that the differential mass function of the L1536 simulation agrees with a reference mass function; here we use as the reference mass function the measurement from the L768 simulation. For mass bins in which the unweighted halo mass function in the L1536 run is below the target mass function, halos are assigned a weight greater than unity. By applying this weight to the halos in the L1536 run, the new, `weighted' mass function agrees with the target mass function exactly by construction. In practice we set the weights to unity above some mass, e.g. $5 \times 10^{13} h^{-1} M_{\odot}$, to avoid being affected by fluctuations at high masses in the mass function measured from the L768 run due to sample variance. The limiting factor which sets the new resolution limit of the weighted halo catalogue is the error that we are prepared to tolerate on the halo clustering. 

With the weighting scheme, the halo correlation function is estimated by including the weight assigned to each halo in the pair count. As a first simple illustration we consider the clustering of samples of main subhalos defined by  different lower mass thresholds in the left and centre-left panels of  Fig.~\ref{fig2:2PCF}. These samples are equivalent to central galaxies in a simple HOD analysis with a sharp transition in the mean occupancy from 0 to 1 central per halo. In each case, the clustering of the halos in the L768 simulation is estimated without applying any weights, i.e. all halos in this case have the same weight of unity, whereas for the L1536 simulation the weights derived from forcing the halo mass function to match that in the L768 run are applied and included in the estimation of the correlation function. Due to the shape of the halo mass function, halos close to the minimum mass that defines each sample contribute importantly to the abundance of halos in the sample and to the clustering.  

%The left-most panel of Fig.~\ref{fig2:2PCF} shows the clustering for a halo mass cut for which the mass functions in the L768 and L1536 runs agree with one another even without applying any weights to the L1536 run. This shows how well the subhalo correlation functions measured from the different simulation volumes agree with one another. The turn over at small pair separations arises because we are measuring the clustering of main subhalo pairs, which means that there is an exclusion effect at small separations. 

The left-most panel of Fig.~\ref{fig2:2PCF} shows the clustering measured for a subhalo sample defined by a  mass cut of $10^{12} h^{-1}M_{\odot}$, close to which modest weights have been applied in the lower resolution run; for halos in which the weight is {\it not} unity the average weight applied in this case is $1.15$. The clustering measured in the L1536 run for this halo sample, after applying the weights, agrees remarkably well with that measured in the L768 run, being within the estimated errors on the correlation function down to $\sim 3 h^{-1} {\rm Mpc}$. In the case without weights, the clustering measured for this halo sample in the lower resolution run is systematically shifted upwards by around 5 per cent compared to that measured in the higher resolution simulation. 
%The shift downwards in the amplitude of the weighted correlation function can be understood in terms of the boost of the contribution of the least biased halos in the sample, those near the mass limit, to the overall clustering signal.   

The centre-left panel of Fig.~\ref{fig2:2PCF} shows the limit of the performance of our weighting scheme. This halo sample is again defined by a lower mass threshold than the one in the left-most panel. For the halos with a weight greater than unity, the average weight in this example is $1.8$. Again, by construction, the weighted sample matches the abundance of halos in the L768 run to better than 1 per cent (the agreement could be further improved by using narrower bins to measure the halo mass function in the mass range where weights greater than unity are derived). The clustering in the weighted sample matches that in the L768 simulation over a reduced range of scales, compared to the other cases, being within the errors down to  $10 h^{-1} {\rm Mpc}$. We note that the clustering of the unweighted halos for this sample is 10 to 15 per cent higher than the `target' measurement from the higher resolution simulation. If this is the error in the clustering that we are prepared to accept, agreement down to  intermediate scales, rising to a 5 per cent excess approaching $\sim 1 h^{-1} {\rm Mpc}$, then the mass resolution of the halo catalogue has been extended down to halos with $\sim 11$ particles. To put this into context, the abundance of the halo samples starts to deviate between the L1536 and L768 simulations at a mass corresponding to around 550 particles. 

As a second example we consider samples that are more comparable to those in HOD analyses, in which the occupation of halos by centrals moves from zero to one per halo more gradually than in the example above. The width of the transition is one of the HOD parameters; $\sigma_{\log M}$. Larger values of $\sigma_{\log M}$ mean that lower mass halos contribute central galaxies to the sample. (Note that we do not consider satellite galaxies in any of our examples; all galaxies within a halo would be assigned the weight of the halo to compute the abundance of galaxies and to estimate their clustering.)   

In the popular five parameter HOD model the mean occupation of halos by centrals depends on the parameters $M_{\rm min}$ and $\sigma_{\log M}$ through (Eqn 1 from \citealt{Zheng:2005}):
\begin{equation}
    \left< N_{\rm cen} \right> = \frac{1}{2}\left[ 1 + \text{erf} \left( \frac{\log M - \log M_{\rm min}}{\sigma_{\log M}} \right) \right]. \label{eq:HOD_cen}
\end{equation}
The centre-right and right panels of  Fig.~\ref{fig2:2PCF} show the correlation functions measured from the L768 and L1536 runs for a fixed value of $M_{\rm min}=4 \times 10^{12} h^{-1} M_{\odot}$, varying $\sigma_{\log M}$. The width of the transition from $\left< N_{\rm cen} \right>=0$ to $\left< N_{\rm cen} \right>=1$ gets broader in mass as the value of $\sigma_{\log M}$ increases. This means that lower mass subhalos are contributing to the correlation function shown in the right-most panel of Fig.~\ref{fig2:2PCF} compared to the centre-right panel. In the centre-right panel of Fig.~\ref{fig2:2PCF}, the transition from all halos being empty to all containing a central is relatively narrow. 
%Given the value considered for $M_{\rm min}$, close to the 100 particle limit for L1536, there is little difference between the weighted and unweighted estimate of the correlation function of this halo sample. The results from L768 agree with both those from the L1536 box. The departures from a ratio of unity show the sample variance fluctuations in the clustering measured from the smaller box, for this number density of halos. 
As $\sigma_{\log M}$ increases, due to the shape of the halo mass function, the number of haloes in the samples increases. The centre-right panel of Fig.~\ref{fig2:2PCF} shows that applying the weighting scheme allows us to recover the correlation function down to $2.5 h^{-1}$Mpc. Again, without applying any weights, the clustering measured in the L1536 box would be systematically shifted upwards by 5 per cent. For the broadest transition considered, with $\sigma_{\log M}= 0.9$, the weighted correlation function is within a few per cent of the estimate from the higher resolution L768 simulation; without weights the estimate is too high by more than 15 per cent. 

\textRed{We have tested the performance of our method at $z=1$. In this case, the marginally resolved halos have a clustering bias that is greater than unity and this poses a challenge to the method. In the simplest case, using a mass threshold of $\log M_{min} = 12.0\ h^{-1}\ M_{\odot}$ to populate haloes with central galaxies, the initial disagreement in the measured clustering is around 6\% on scales larger than $1\ h^{-1}$ Mpc. After applying  the weighting scheme, this disagreement drops to 3\%. The performance of the scheme is less good than at $z=0$, but still represents an improvement over doing nothing. The situation is similar for the case with $\log M_{min} = 12.6\ h^{-1}\ M_{\odot}$ and $\sigma_{\log M}=0.5$. Here the difference in the correlation measured from the simulations without weighting differs by 8\% on scales $1 < r/ \text{Mpc}\ h^{-1} < 20$. When we apply the weighting scheme, the discrepancy more than halves to a 3\% of disagreement. In these two cases we are applying weights with values between 5 and 10 to haloes with around 20 particles.}

We end by investigating the incomplete or `partially' resolved halo population in the lower  resolution L1536 simulation. What is special about the subhalos that are picked up by FoF and {\sc subfind}, at masses for which the subhalo samples in this run are incomplete? We address this by measuring the local environment around halos as a function of mass, by measuring the distribution of counts-in-cells centred on halos, and comparing the measurements between the L1536 and L768 runs. We use cubical cells of side $1.6 h^{-1}$Mpc which sample the density field defined by the dark matter particles. We find that the counts-in-cells distributions around subhalos that are well resolved in each simulation are essentially the same. The difference in shot noise (mean particle density) does not affect the count distributions because centring on a halo biases the counts to high densities. The left and centre panels of Fig.~\ref{fig3:density} contrast the cell count distributions measured in the two simulations in mass bins for which the mass functions are different. As the mass bin shifts to lower masses, the difference between the local environments of the subhalos that are identified changes, with marginally resolved subhalos from the L1536 run tending to be found in higher density environments than the true distribution, according to the measurement from the L768 run. The right panel of  Fig.~\ref{fig3:density} shows the difference in the local density around subhalos for a sample with a sharp mass cut at $5.6 \times 10^{11} h^{-1} {\rm M_{\odot}}$. The mass range shown is that for which greater than unity weights are applied in our scheme. The unweighted cell-count distribution is shown by the solid green line; the weighted distribution, using the weights computed in 10 mass bins, is shown by the green dashed line and is remarkably close to the distribution found in the higher resolution L768 run.        

%Here we investigate why the clustering in the incomplete or partially resolved samples is different from that in the high res simulations in which all of these halos are resolved. There is an offset in the local density of the halos that are picked up in the low resolution sim, compared to the same mass halos in the high resolution sim. 

%\begin{figure}
%    \centering
%    \includegraphics[width=0.9\columnwidth]{figures_pdf/N_Vnormed_density_L768_1536_weighted_Mbin_11.75_13.25.pdf}
%    \caption{The distribution of counts in cells centred on subhalos in the broad mass range between $11.75 \leq \log(M/ M_{\odot} h^{-1}) < 13.25$, measured from the L768 (red) and L1536 (green) simulations. We count the haloes using the weights defined by our scheme to recreate the density distribution of the L768 simulation (dashed-line green).}
%    \label{fig3:density_weighted}
%\end{figure}

\section{Summary and Conclusions} \label{sec:conclusions}

Often the mass resolution limit of a simulated halo catalogue is presented as a suspiciously round number, 100+ particles, that may once have been checked but has long since passed into simulation folklore and has become an unquestioned rule of thumb. We have argued that for some studies, for example simple clustering analyses, such limits are overly conservative as we are not interested in quantities that are more difficult to calculate, such as the internal structure of the halo. We have gone a step further and presented a simple weighting scheme to compensate for `missing' halos by upweighting those that are recovered by the halo finder. Our scheme is able, by construction, to reproduce a `target' number density of halos, and returns improved estimates for the clustering of halo samples. Depending on one's error tolerance for the accuracy of the clustering predictions, we showed an example in which this scheme extended the mass resolution of a halo catalogue down to objects made of 11 particles.  

As presented, our scheme requires at least two simulations. One is designated as the high resolution simulation and sets the target or benchmark for the halo sample statistics. This simulation is used to provide the `correct' answer for the halo mass function, and to provide some indication of the expected clustering for different halo samples. No weights are applied to the halos in the high resolution simulation. The second simulation is lower resolution, typically because it models the growth of structure in a much larger volume, with a similar or reduced number of particles than the high resolution simulation. The purpose of this simulation could be to access clustering predictions on larger scales than could be reached with the high resolution simulation, such as the scale of the baryonic acoustic oscillations. Also, many copies of the low resolution simulation could be run using an approximate simulation method to generate many realisations of halo samples for error estimation. Examples of both these use cases can be found in \cite{Cesar:2021}. By extending the usable halo catalogue derived from the low resolution run down to lower masses, significant computational resources can be saved. 

The subhalo finding algorithm recovers a fraction of the expected halos in the mass range that is considered `sub-resolution'. We showed that these objects have higher local overdensities than halos in the same mass range that are fully resolved in a higher resolution simulation. The details of which halos are found will no doubt depend somewhat on the subhalo finder algorithm used, and perhaps on the simulation code itself. \textRed{Our scheme does not assign weights using any spatial information, and so cannot ``correct" the clustering for halos in a single mass bin. Our approach works for samples defined by a mass threshold, for which there are several bins in the mass function from which the halos acquire different weights greater than unity.}

The scheme that we have proposed allows the resolution of a halo catalogue to be extended down to small particle numbers by applying a correction to the halos that we do see to account for those that we do not find. Ultimately, the scheme breaks down at the halo mass for which the errors in the clustering prediction become unacceptable. This approach is therefore different to those that try to account for assembly bias in marginally resolved halos (e.g. \citealt{Ramakrishnan:2021}). Assembly bias which arises when the clustering of halos in a given mass range also depends on an internal property, such as formation time, environment or concentration \cite{Gao:2007}. \cite{Ramakrishnan:2021} attempt to estimate internal halo properties from marginally resolved halos (e.g. with 30 particles) in order to build mock catalogues which include assembly bias. In principle it should be possible to combine the two approaches to build more accurate mock catalogues.

\section*{Acknowledgements}
We acknowledge conversations with Idit Zehavi and Sergio Contreras. 
JA acknowledges support from CONICYT PFCHA/DOCTORADO BECAS CHILE/2018 - 72190634. CMB and PN acknowledge support from the Science Technology Facilities Council through ST/T000244/1. NP acknowledges support from Fondecyt Regular project 1191813 and BASAL CATA AFB-170002. We acknowledge  financial support from the European Union's Horizon 2020 Research and Innovation programme under the Marie Sklodowska-Curie grant agreement number 734374 - Project acronym: LACEGAL
This work used the DiRAC@Durham facility managed by the Institute for Computational Cosmology on behalf of the STFC DiRAC HPC Facility (www.dirac.ac.uk). The equipment was funded by BEIS capital funding via STFC capital grants ST/K00042X/1, ST/P002293/1, ST/R002371/1 and ST/S002502/1, Durham University and STFC operations grant ST/R000832/1. DiRAC is part of the National e-Infrastructure.
%%%%%%%%%%%%%%%%%%%%%%%%%%%%%%%%%%%%%%%%%%%%%%%%%%
\section*{Data Availability}
The halo mass functions and correlation functions generated in this paper are available on reasonable request. To access the subhalo catalogues, the authors of the papers describing the simulations should be contacted



\bibliographystyle{mnras}
\bibliography{sideshow} % if your bibtex file is called example.bib

%\begin{thebibliography}{99}
%\end{thebibliography}

%%%%%%%%%%%%%%%%%%%%%%%%%%%%%%%%%%%%%%%%%%%%%%%%%%

%%%%%%%%%%%%%%%%% APPENDICES %%%%%%%%%%%%%%%%%%%%%

%\appendix



%%%%%%%%%%%%%%%%%%%%%%%%%%%%%%%%%%%%%%%%%%%%%%%%%%

% Don't change these lines
\bsp	% typesetting comment
\label{lastpage}
\end{document}